\documentclass[final,5p,times,twocolumn]{elsarticle}

\biboptions{sort&compress}

\usepackage{lineno,hyperref}
\usepackage{graphicx}
\usepackage{latexsym}
\usepackage{textcomp}
\usepackage{amssymb}
\usepackage{amsbsy}
\usepackage{amsmath}
\usepackage{bm}
\usepackage{longtable}
\usepackage{subfigure}
\usepackage{epsfig}
\usepackage{dcolumn}
\usepackage[Euler]{upgreek}
\usepackage{multirow}
\usepackage{footnote}
\usepackage{tabulary}
\usepackage{color}
\usepackage{xcolor}

\DeclareMathAlphabet{\mathsf}{OT1}{phv}{b}{n}

\newcommand{\Vector}[1]{\ensuremath{\mathbf{#1}}}

\newcommand{\crossVorg}{\ensuremath{%
         \setbox0=\hbox{$V$}
        V \kern-\wd0{\raise.3ex\hbox{$\relbar$}}}}

\newcommand{\crossVxx}[2]{%
	{\setbox0=\hbox{$#1#2V$}
         \setbox1=\hbox{$#1#2$}
         \setbox2=\hbox{$#1V$}
         \dimen1=\wd0
	 \advance\dimen1-\wd1
         \raise.2\ht0\hbox{$#1#2$}\kern-.4\wd0}}

\newcommand{\degree}{\ensuremath{^\circ}}
\newcommand{\degC}{\ensuremath{\degree\mathrm{C}}}

\usepackage{pifont}

\newcommand{\Rhth}{\ensuremath{R_{\mathrm{H}}^{\theta}}}

\newcommand{\Rv}{\ensuremath{R_{\eta}}}

\newcommand{\Na}{\ensuremath{N_{\mathrm{A}}}}

\newcommand{\Rh}{\ensuremath{R_{\mathrm{H}}}}

\newcommand{\ag}{\ensuremath{\alpha_{\mathrm{G}}}}
\newcommand{\ah}{\ensuremath{\alpha_{\mathrm{H}}}}

\newcommand{\aeta}{\ensuremath{\alpha_{\mathrm{\eta}}}}

\newcommand{\Uer}{\ensuremath{U_{\mathrm{\eta R}}}}
\newcommand{\etas}{\ensuremath{\eta_{{s}}}}
\newcommand{\etap}{\ensuremath{\eta_{{p}}}}

\newcommand{\etapo}{\ensuremath{\eta_{\mathrm{p,0}}}}

\newcommand{\leta}{\ensuremath{\lambda_{\mathrm{\eta}}}}

\newcommand{\ivisc}{\ensuremath{[\eta]}}

\newcommand{\ccs}{\ensuremath{c / c^{*}}}

\journal{Current Opinion in Colloid \& Interface Science}

\begin{document}

\begin{frontmatter}

\title{Universal dynamics of dilute and semidilute solutions of flexible linear polymers}

\author{J. Ravi Prakash}
\address{Department of Chemical Engineering, Monash University, Melbourne, VIC 3800, Australia}
\ead{ravi.jagadeeshan@monash.edu}
\ead[url]{http://users.monash.edu.au/~rprakash}

\begin{abstract}
\noindent Experimental observations and computer simulations have recently revealed universal aspects of polymer solution behaviour that were previously unknown. This progress has been made possible due to developments in experimental methodologies and simulation techniques, and the formulation of scaling arguments that have identified the proper scaling variables with which to achieve data collapse, both for equilibrium and nonequilibrium properties. In this review, these advances in our understanding of universal polymer solution behaviour are discussed, along with a brief overview of prior experimental observations and theoretical predictions in order to provide a context for the discussion. Properties for which the existence of universality is still unclear, and those for which it has not yet been verified theoretically are highlighted, and theoretical predictions of universal behaviour that require experimental validation are pointed out.  
\end{abstract}


\end{frontmatter}

\section{Introduction}

Solutions of flexible linear neutral polymers have several properties that exhibit universal behaviour at large length and time scales, independent of the local structure of the polymer molecule, such as the  chemistry of the monomer. This striking feature of polymer solution behaviour is observed experimentally across a wide range of static and dynamic properties at equilibrium, and also for non-equilibrium rheological properties. For instance, the increase in the equilibrium size of a polymer coil with temperature and molecular weight is independent of the details of the polymer and solvent when the data is represented in terms of appropriately chosen variables. In shear flow, the shear thinning of viscosity observed in different polymer-solvent systems can be collapsed on to a master plot when interpreted in terms of suitably rescaled variables. The existence of universality has provided the motivation and framework for the development of theoretical models which underpin our current understanding of polymer solution behaviour, and many of the experimental observations have been validated and successfully explained by these models~\cite{degennes,Doi1986,cloizeaux1990,schafer,Rubinstein2003}. In this article, some additional and novel aspects of the universal behaviour of polymer solutions that have been revealed and validated in the past few years due to developments in experimental methodologies and simulation techniques are reviewed. Further, experimental observations that are yet to be modelled theoretically and simulation predictions of universal behaviour that require experimental validation are highlighted. 
 
The origin of universality lies in the scale-invariant character of high molecular weight polymers, which follows from their conformations looking the same on a large range of length scales~\cite{degennes,Rubinstein2003}. The existence of scale invariance has many implications for the development of theoretical models~\cite{Duenweg2018}. Firstly, it implies that the dependence of properties of polymer solutions on physical quantities can be described in terms of scaling laws. These laws suggest that rather than explicitly depending on the temperature ($T$), the monomer concentration ($c$), the molecular weight ($M$), and strain rate ($\dot \gamma$ in shear flow and $\dot \epsilon$ in extensional flow), polymer solution properties depend on appropriately scaled nondimensional combinations of these variables. Secondly, since scale invariance implies independence from chain microstructure, it is sufficient to use simple coarse-grained models for the polymer (such as the bead-spring chain model), that discard all details of the chemistry of the monomer, but only retain instead its essential physical aspects. In the case of linear, neutral polymers that are not sufficiently long to be entangled, experimental observations are successfully captured by coarse-grained models that include (i) the \emph{connectivity} of monomers along the chain backbone~\cite{rouse53,birdetal2}, (ii) the effective \emph{excluded volume interaction} between pairs of monomers, which is a consequence of monomer-monomer, monomer-solvent, and solvent-solvent enthalpic interactions~\cite{Flory1953,edwards65,schafer}, and (iii) the phenomenon of \emph{hydrodynamic interactions} that describes the pair-wise coupling of the motions of different monomers due to the propagation of momentum by the solvent~\cite{zimm56,degennes,birdetal2,RaviBookchapter1999}.  

The inclusion of nonlinear phenomena such as excluded volume and hydrodynamic interactions into molecular theories makes them impossible to solve exactly analytically. Theoretical approaches inevitably introduce various approximations in order to obtain a tractable model, and as a consequence, a large majority of theoretical predictions of polymer solution behaviour are approximate in nature~\cite{RaviBookchapter1999,Wang2017}. The excellent agreement of theoretical predictions with experimental observations in many instances, particularly for equilibrium and near-equilibrium properties, indicates that in these cases at least, these approximations are not severe and indeed justified.

Computer simulations, on the other hand, enable the exact solution of the underlying polymer model (within statistical and computational error), and as a consequence  allow a more thorough exploration of the success or failure of the model. While universal static properties of polymer solutions, which only require the treatment of excluded volume interactions, have been extensively studied by computer simulations~\cite{Duenweg2018}, universal equilibrium and non-equilibrium dynamic properties have not been studied as widely.  This is because they require the incorporation of hydrodynamic interactions, which is challenging within the framework of methodologies conventionally used to simulate static properties. Recently, however, there has been significant progress in the development of simulation algorithms that account for hydrodynamic interactions, and they have been used to predict both dynamic properties at equilibrium and far-from-equilibrium rheological properties~\cite{BD1999,Gompper2009,Duenweg2009,Ottinger1996,Huang2010,Stoltz2006,Duenweg2018,Saadat2016,Fiore2017,Dyer2017}. In this review I will discuss attempts in the past few years at applying these new algorithms to investigate the \emph{universal} aspects of the dynamics of polymer solutions. In particular, I will focus on Brownian dynamics simulations, since many of the recent results have been obtained with this technique. More comprehensive reviews of progress in the description of the general rheological behaviour of polymer solutions can be found in the excellent articles by Larson~\cite{Larson2005}, Shaqfeh~\cite{Shaqfeh2005} and Schroeder~\cite{Schroeder2018}.

The plan of the paper is as follows. In Section~\ref{sec:equil}, universal behaviour at and close to equilibrium is considered. Sections~\ref{sec:thermal} and~\ref{sec:conc} discuss the appropriate scaling variables that must be used to describe the the thermal crossover from $\theta$ to perfectly good (athermal) solvents, and the concentration crossover from dilute to concentrated solutions, respectively. A brief summary of prior experimental observations and theoretical predictions of universal static and dynamic properties is given, along with an account of some recent developments. Section~\ref{sec:double} discusses the semidilute concentration regime, with the solvent quality lying between $\theta$  and athermal solvents, where several observations and predictions of universal behaviour have been made recently. Nonequilibrium behaviour is discussed in Section~\ref{sec:nonequil}, with  shear flow considered in Section~\ref{sec:shear}  and extensional flow in Section~\ref{sec:ext}. Though experimental measurements of the flow of synthetic polymer solutions have been carried out for several decades, the recent use of DNA as a model polymer has led to the reporting of new observations, and attempts to interpret data within the framework of universal behaviour. Section~\ref{sec:dilshear} on dilute solutions emphasises the large gap that currently exists between experiments and theory with regard to several aspects of the observed behaviour in shear flow, and highlights the problems with establishing the existence of universal behaviour. Both simulation predictions and bulk and single molecule experimental observations of extensional flow are discussed in Section~\ref{sec:ext}, with simulations suggesting the existence of universal behaviour that is yet to be demonstrated experimentally.  Section~\ref{sec:conclu} summarises the key recent results, and suggests some future directions.

\section{Universal behaviour at and close to equilibrium \label{sec:equil}}

\subsection{Thermal crossover in dilute solutions \label{sec:thermal}}

The existence of scale invariance implies that all properties are governed by power laws~\cite{Duenweg2018}.  For instance, the mean size of the chain $R$ does not depend on the precise number of segments $N$ of size $b$ that the chain is subdivided into, provided $b$ is large compared to the size of the monomer, and small compared to $R$. Thus at the $\theta$-temperature $T_\theta$, where monomer-monomer repulsion is balanced by monomer-solvent attraction, the chain size in a dilute solution obeys a power law that describes a random walk (RW), $R_\theta = b N^{1/2} $, while in an athermal solvent, where the energy of monomer-monomer excluded volume interactions is greater than the thermal energy $k_\text{B} T$, the chain obeys self-avoiding walk (SAW) statistics, with size $R_\text{F} = b N^{\nu}$. Here, $k_\text{B}$ is Boltzmann's constant. The Flory exponent $\nu$ has been calculated by renormalisation group theory to have a value of 0.588~\cite{cloizeaux1990,schafer}. Notably, both the exponent values of $\frac{1}{2}$ and $\nu$ are universal for polymer-solvent systems under $\theta$ and athermal solvent conditions, respectively. Further, various average measures of the polymer coil size, such as the end-to-end vector $R_E$, or the radius of gyration $R_G$, all scale identically with $N$. This has a consequence that ratios such as $R_E/R_G$ have universal values, typically denoted as amplitude ratios. For instance, for RW chains, the universal amplitude ratio $R_E/R_G = \sqrt{6}$.

Several computational studies have validated the predictions of universal static properties in perfectly good solvents made by scaling and renormalisation group theories, by using Monte Carlo simulations, with polymer chains represented as self avoiding walks on lattices~\cite{Baschnagel2004,Hsu2011,Landau2014}. A recent extremely fast implementation by Clisby~\cite{Clisby2010} of Monte Carlo trial moves in the pivot algorithm, has led to the following predictions for three dimensional SAWs, which are probably the most accurate currently available in the literature~\cite{Clisby2016}: $\nu = 0.58759700(40)$, and $(R_E/R_G)^2 = 6.253531(10)$. 

In general, however, polymer solutions are at temperatures $T$ that lie between $T_\theta$ and the athermal limit. Under these conditions, the existence of self-similarity implies that a type of universal scaling still exists~\cite{degennes,schafer,Rubinstein2003}. In a dilute solution, at temperatures $T \!\!> \!\!T_\theta$, polymer coils swell due to the improvement in solvent quality, and scale invariance implies~\cite{schafer}
\begin{equation}
 R =  R_\theta \, f(z)
 \label{swell}
\end{equation}
where 
\begin{equation}
z = k \left(1 - \frac{T_\theta}{T} \right) N^{1/2}
\end{equation}
is the solvent quality parameter, with $k$ being a polymer-solvent chemistry dependent constant, and $f(z)$ is a \emph{crossover} function that describes the chain's crossover from RW statistics at $T_\theta$ to SAW statistics in an athermal solvent. The form of the function $f(z)$ in the entire crossover region has been derived by renormalisation group calculations~\cite{schafer}, and it has the limiting behaviour, $f(z)  \approx 1$ for $z \le 1$ and $f(z) \sim z^{2\nu-1}$ for $z \gg 1$, consistent with $\theta$ and athermal solvent conditions, respectively.
 
The thermal crossover between $\theta$ and athermal solvents in a dilute polymer solution can be represented visually by imagining that the polymer chain breaks up into a sequence of \emph{thermal blobs}, which are spherical ``blobs" of size $\xi_T$, within which polymer segments consisting of $g_T$ monomers experience a net excluded volume interaction of order $k_\text{B} T$~\cite{degennes,Rubinstein2003}. Since the excluded volume energy within a blob is less than the energy of thermal fluctuations, the chain configuration remains a random walk  below the length scale $\xi_T$. At larger length scales, the thermal blobs exclude each other and exhibit SAW statistics. In the blob framework, the solvent quality parameter can be represented by 
\begin{equation}
 z  = \frac{R_\theta}{\xi_T} 
\end{equation}
with the number of thermal blobs given by $\mathcal{N}_T = N/g_T = z^{2}$~\cite{Jain2012a}. Under $\theta$ conditions the polymer obeys RW statistics, so the entire chain lies within a blob, i.e., $\xi_T \sim R_\theta$, which implies, $g_T \sim \mathcal{O}(N), \mathcal{N}_T \sim  \mathcal{O}(1)$, and $z \le 1$. Under athermal conditions, the entire chain obeys SAW statistics, so $\xi_T \sim b$, which implies, $g_T \sim \mathcal{O}(1), \mathcal{N}_T \sim  \mathcal{O}(N)$, and $z \gg 1$. The solvent quality parameter $z$ is consequently a scaling variable that combines the dependence of solution properties on the molecular weight $M$, and temperature, $T$, and describes the shrinking of the thermal blob size $\xi_T$ from $\mathcal{O}(R_\theta)$ to $\mathcal{O}(b)$ in the crossover regime.

Within the blob ansatz, universality arises because polymers in different polymer-solvent systems look the same on length scales large compared to the blob size,  i.e., the details of the chemistry are buried within a blob. For instance, one can understand the universal swelling of the polymer chain size observed experimentally~\cite{MiyFuj81} as follows. 
Consider two polymer-solvent systems having the same solvent quality, i.e., $z_1=z_2$, which implies the ratio of the chain size under $\theta$-conditions to the thermal blob size is the same in both systems. Since, $R= \xi_T  \mathcal{N}_T^\nu = (R_\theta/z) \, z^{2\nu}$, it follows that the swelling of a chain in both polymer-solvent systems, $R/R_\theta = z^{2\nu-1}$, is identical. This is also immediately obvious from Eq.~(\ref{swell}).

The thermal crossover in dilute polymer solutions has been successfully described by Brownian dynamics simulations, in which the polymer is represented by a coarse-grained bead-spring chain model, with each chain consisting of a sequence of $N$ beads (which act as centers of hydrodynamic resistance) connected by $N - 1$ massless springs that represent the entropic force between two adjacent beads. The time evolution of the nondimensional position, $\Vector{r}_{\nu}$ of a typical bead $\nu$, is governed by a stochastic differential equation, which is integrated numerically~\cite{Ottinger1996}. Within this framework, Prakash and coworkers have shown that the solvent quality can be conveniently controlled with the help of the narrow Gaussian potential~\cite{RaviExclu,Kumar}
\begin{equation}
\label{exvol}
E(\Vector{r}_{\nu \mu}) = \left(\frac{z^*}{{d^*}^3}\right) \exp \left\lbrace -\frac{1}{2}\frac{{\Vector{r}_{\nu \mu}}^2}{{d^*}^2} \right\rbrace 
\end{equation}
which determines the force due to excluded volume interactions between any two beads $\mu$ and $\nu$. Here, $z^*$ is the strength of the excluded volume interactions, $d^*$ is the range of the interaction, and $\Vector{r}_{\nu \mu} = \Vector{r}_{\nu} - \Vector{r}_{\mu}$ is the connector vector between beads $\mu$ and $\nu$. The narrow Gaussian potential is a means of regularizing the Dirac delta potential since it reduces to a $\delta$-function potential in the limit of $d^*$ tending to zero. A mapping between experiments and simulations is achieved by setting $z=z^* \sqrt{N}$, with $z^*$ being a measure of the departure from the $\theta$-temperature~\cite{Kumar,Sunthar06}. As a result, for any choice of $N$, $z^*$ is chosen to be equal to $z/\sqrt{N}$ such that the simulations correspond to the given experimental value of $z$. For reasons elaborated in Refs.~\cite{prakash2001influence,Kumar}, the parameter $d^*$ is irrelevant for sufficiently long chains. 

Universal predictions, independent of details of the coarse-grained model used to represent a polymer, are obtained in the limit of long chains, since the self-similar character of real polymer molecules is captured in this limit. In Brownian dynamics simulations, predictions in the long chain limit are obtained by accumulating data for finite chain lengths and extrapolating to the limit $N \to \infty$. This procedure has been used successfully by Kumar and Prakash~\cite{Kumar} to calculate the  universal crossover behaviour of the swelling ratio of the radius of gyration $\ag (z) = R_G / R_G^\theta$, and shown to be in excellent agreement with the experimental measurements of Miyaki and Fujita~\cite{MiyFuj81}. A similar extrapolation methodology was used by Kr\"{o}ger et~al.~\cite{Kroger2000} to obtain predictions for a number of universal amplitude ratios under $\theta$-conditions, which they compared with experimental values reported previously. 

\begin{figure}[tbp]
\begin{center}
\resizebox{\linewidth}{!}{\includegraphics*{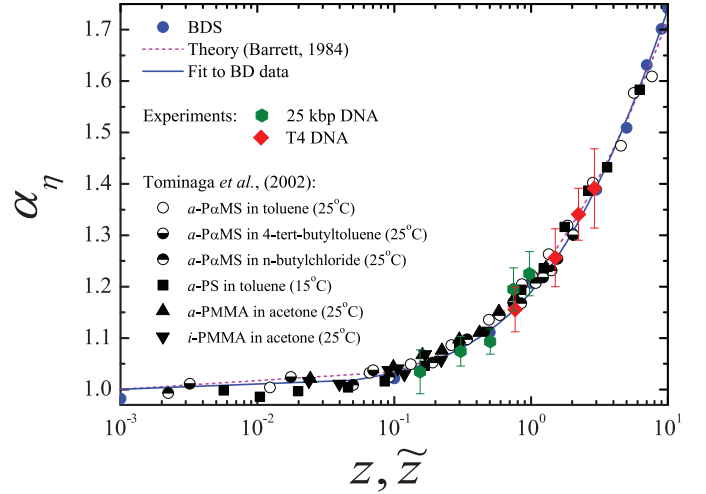}}
\end{center}
\vskip-20pt
\caption{Crossover swelling of the viscosity radius from $\theta$ to good solvents. For wormlike chains, the solvent quality parameter $z$ is modified to $\tilde z$ to account for chain stiffness~\cite{yamakawa1997}. Experimental measurements of the swelling of 25 kbp and T4 DNA  are represented by the filled hexagons and diamonds, respectively, while the remaining symbols represent data on various synthetic wormlike polymer-solvent systems collated in Ref.~\cite{Tominaga20021381} The filled blue circles are the predictions of the current BD simulations. The solid line represents a fit to the Brownian dynamics data, while the dotted red line is the prediction of the  Barrett equation~\cite{Barrett1984} for \aeta. Reproduced with permission from Ref.~\cite{Pan2014b}. }
\label{fig:alphaeta}
\vskip-10pt
\end{figure}

Two other size ratios that are measured frequently are $\ah = \Rh / \Rhth$ (where \Rh, the hydrodynamic radius, is proportional to the inverse of the chain's diffusion coefficient), and $\alpha_{\eta} = {R_{\eta}}/{R_{\eta}^{\theta}}=\left({[\eta]_0}/{[\eta]_{0}^{\theta}}\right)^{{1}/{3}}$, where $R_{\eta}$ is the viscosity radius, defined by the expression,
\begin{equation}
\label{eq:RV}
R_{\eta} \equiv \left(\frac{3 [\eta]_0 M }{10 \pi N_{A}}\right)^{{1}/{3}}
\end{equation}
with \Na\ being the Avogadro's constant, and $\ivisc_0$ the zero shear rate intrinsic viscosity. Several experimental studies~\cite{MiyFuj81,Tominaga20021381,Hayward19993502} have established that both these ratios can be collapsed onto master plots for different polymer-solvent systems, in the crossover regime. Notably, however, the universal curve for $\ag (z)$ (which is a ratio of static properties), is significantly different from the universal curves for $\ah (z)$ and $\aeta (z)$, which are ratios of dynamic properties. There have been many attempts to understand the origin of this difference in crossover behaviour, and to predict analytically and numerically, the observed universal curves (see Ref.~\cite{Jamieson2010} for a recent review). 
Since both the hydrodynamic and viscosity radii are dynamic properties, hydrodynamic interactions play a crucial role in determining their swelling functions \ah\ and \aeta. By incorporating hydrodynamic interactions with a Rotne-Prager-Yamakawa tensor, and extrapolating finite bead-spring chain data from Brownian dynamics simulations to the long chain limit, Sunthar and Prakash~\cite{Sunthar06} have shown that the difference between \ag\ and \ah\ is in fact due to the presence of fluctuating hydrodynamic interactions in the non-draining limit. The use of the narrow Gaussian potential, and accounting for fluctuating hydrodynamic interactions, consequently enables quantitatively accurate predictions of \ag\ and \ah\ as functions of $z$~\cite{Kumar,Sunthar06}.

The problem of predicting the crossover function $\aeta (z)$ was addressed recently by Pan et al.~\cite{Pan2014b} by adopting essentially the same procedure, namely extrapolating finite bead-spring chain data, obtained by Brownian dynamics with hydrodynamic interactions incorporated, to the long chain limit. They also examined the prediction of the crossover function for the universal amplitude ratio $\Uer (z)$, defined by
\begin{equation}
\label{eq:uetar}
U_{\eta R} \equiv \frac{5}{2} \left( \frac{\Rv}{R_G} \right)^{3} = \frac{6^{\frac{3}{2}}}{(4\pi/3)}\, \frac{\Phi}{\Na} 
\end{equation}
$\Uer (z)$ is related to the well known Flory-Fox constant $\Phi$ as shown in Eq.~(\ref{eq:uetar}) above, where $\Phi =  [\eta]_0 M /6^{\frac{3}{2}} R_G^{3}$ ~\cite{Rubinstein2003}. In addition to Brownian dynamics simulations, Pan et al.~\cite{Pan2014b} have carried out systematic measurements of the intrinsic viscosity of two different molecular weight samples of linear double-stranded DNA at a range of temperatures in the presence of excess salt, and examined the crossover scaling of \aeta\ and $\Uer (z)$. Even though a number of experimental measurements of the Flory-Fox constant under good solvent conditions have been reported in the literature~\cite{Jamieson2010}, the behaviour of $\Phi$ with varying solvent conditions and molecular weight had not been understood with any great certainty prior to this work. The results of their simulations and experimental measurements of \aeta\ are displayed in Fig.~\ref{fig:alphaeta}, along with earlier experiments on synthetic polymer solutions, and the analytical prediction by the Barrett equation~\cite{Barrett1984}. The excellent comparison with earlier observations of the behaviour of synthetic polymers establishes that DNA solutions (in the presence of excess salt) exhibit universal scaling, and verifies the earlier estimate of the $\theta$-temperature and solvent quality of DNA solutions, using static and dynamic light scattering, by Pan et al.~\cite{Pan2014a}. The close agreement between the predictions of Brownian dynamics simulation results and experiments shows that  by including fluctuating excluded volume and hydrodynamic interactions, quantitatively accurate prediction of the crossover scaling of \aeta\ can be obtained, free from the choice of arbitrary model parameters. Further, as in the case of \ag\ and \ah, the difference between the crossover scaling of \ag\ and \aeta\ is shown to arise from the influence of hydrodynamic interactions in the non-draining limit.

A striking success of Zimm theory~\cite{zimm56}, which treats hydrodynamic interactions in an approximate pre-averaged way, is its quantitatively accurate description of the frequency dependence of the small amplitude oscillatory material functions $G^\prime$ and $G^{\prime\prime}$, under $\theta$ conditions~\cite{birdetal2}. Indeed improvements in closure approximations used to treat hydrodynamic interactions do not lead to any improvement in the prediction of these linear viscoelastic properties~\cite{RaviBookchapter1999}. Lodge and coworkers have shown that $G^\prime$ and $G^{\prime\prime}$ exhibit universal behaviour when suitably normalised, and have carefully examined the influence of solvent quality on the power law scaling of  $G^\prime$ and $G^{\prime\prime}$ with the frequency of oscillations, $\omega$~\cite{Sahouani1992,Larson1999}. By comparing various analytical approaches that incorporate excluded volume effects, they conclude that the results of renormalisation group calculations lead to the most consistent comparison with experimental data. While there has been some effort at using Brownian dynamics simulations to predict the frequency dependence of $G^\prime$ and $G^{\prime\prime}$ under $\theta$-solvent conditions~\cite{Cruz2012}, there have been no attempts so far to study solvent quality effects on small amplitude oscillatory material functions within the modern framework of Brownian dynamics simulations discussed above.

\subsection{Concentration crossover \label{sec:conc}}

With increasing monomer concentration, polymer coils that are isolated from each other in the dilute limit, begin to interact with each other hydrodynamically, until at the overlap concentration $c^*$, they just begin to touch each other and fill space. At this point, since the concentration of monomers within each coil is the same as the overall monomer concentration, $c=c^*=N/R^3$, and since $R=bN^\nu$, we have
\begin{equation}
\label{eq:cstar} 
c^* = b^{-3} N^{-(3\nu-1)}
\end{equation}
For concentrations greater than $c^*$ in a perfectly good solvent, the packing of swollen polymer coils (obeying SAW statistics) leads to the screening of excluded volume interactions (the so-called Flory screening). At sufficiently high concentrations, excluded volume interactions are completely screened, and polymer chains follow RW statistics (as is also the case in dense unentangled polymer melts). The gradual screening of excluded volume interactions  with increasing concentration, which leads to a  crossover in which chains obeying SAW statistics begin to obey RW statistics, can be intuitively understood with the concept of a correlation blob~\cite{degennes,Rubinstein2003}. The picture of a polymer solution that underpins this concept consists essentially of three elements: (i) The correlation blob sets a length scale below which segments on a polymer chain are unaware of other polymers in the system. As a result, they behave as though they are in a dilute solution, and obey SAW statistics. If $\xi_c$ is the size of a correlation blob, and there are $g_c$ monomers in it, it follows that
\begin{equation}
\label{eq:xic} 
\xi_c = b g_c^{\nu}
\end{equation}
(ii) For concentrations greater than $c^*$, polymer chains can be visualised as a sequence of correlation blobs, with the system as a whole consisting of space filling correlation blobs, i.e., it is a melt of correlation blobs. The concentration of monomers within each blob is then the same as the overall monomer concentration,
\begin{equation}
\label{eq:cinblob} 
c = \frac{g_c}{\xi_c^{3}}
\end{equation}
(iii) Excluded volume interactions are completely screened on length scales larger than that of the correlation blob. As a result, a polymer chain is a RW of correlation blobs, and its size is given by
\begin{equation}
\label{eq:semidiluteR} 
R = \xi_c \left(\frac{N}{g_c}\right)^{1/2}
\end{equation}
By combining Eqs.~(\ref{eq:xic}) to~(\ref{eq:semidiluteR}), it is straight forward to show that the size of individual chains obeys the following power law~\cite{degennes,Rubinstein2003}
 \begin{equation}
 \label{eq:Rsemidilute}
 R =  R_\text{F} \,  \left(\frac{c}{c^*} \right)^{-\tfrac{2\nu-1}{2(3\nu-1)}} 
\end{equation}
At a threshold concentration $c = c^{**}$, the size of a correlation blob, which is decreasing with increasing concentration, becomes of order of monomer size. The term \emph{semidilute} is used to describe the range of concentrations originating at $c=c^*$, and extending until $c=c^{**}$, and as a consequence, it encompasses the concentration crossover from a dilute solution (where chains obeys SAW statistics), to the unentangled concentrated limit (above $c^{**}$), where chains obeys RW statistics. 

Equation~(\ref{eq:Rsemidilute}) suggests that the natural scaling variable in the semidilute regime is $(c/c^*)$. In terms of this scaling variable, one can show that the size of a correlation blob is given by
\[ \xi_c =  R_F \left(\frac{c}{c^*}\right)^{-\tfrac{\nu}{3\nu-1}} \]
and the number of correlations blobs on a chain,  $\mathcal{N}_c = N/g_c$, is given by the expression
\[ \mathcal{N}_c =  \left(\frac{c}{c^*}\right)^{\tfrac{1}{3\nu-1}} \]
At $c = c^{*}$, the entire chain lies within a correlation blob, i.e., $\xi_c \sim R_F$, with $g_c \sim \mathcal{O}(N)$, and $\mathcal{N}_c \sim  \mathcal{O}(1)$. For $c \gtrsim c^{**}$, the entire chain obeys RW statistics, with $\xi_c \sim b$, $g_c \sim \mathcal{O}(1)$, and $\mathcal{N}_c \sim  \mathcal{O}(N)$. The scaling variable $(c/c^*)$ consequently combines the dependence of solution properties on the molecular weight $M$, and concentration, $c$, and describes the shrinking of the correlation blob size $\xi_c$ from $\mathcal{O}(R_F)$ to $\mathcal{O}(b)$ in the concentration crossover regime.

The screening of excluded volume interactions in the semidilute regime is accompanied by a simultaneous screening of hydrodynamic interactions~\cite{Edwards1974,DeGennes1976bw,degennes,Edwards1984,Ahlrichs2001,Rubinstein2003,Sing2018}. Essentially, correlations in the motion of different chain segments, that arise from hydrodynamic interactions, fade away with increasing concentration because they are randomised due to chain-chain collisions. In other words (at long times; see below), the velocity perturbation due to a point force doesn't decay inversely with distance $r$ from the force, but rather as $(1/r)\exp(-r/\xi_H)$, where $\xi_H$ is the hydrodynamic screening length. De Gennes~\cite{DeGennes1976bw} argued that the length scale associated with hydrodynamic screening is identical to that for the screening of excluded volume interactions, namely $\xi_H \sim \xi_c$. Since the Zimm model~\cite{zimm56} accounts for hydrodynamic interactions, while the Rouse model~\cite{rouse53} does not, it implies that in this ansatz, chain segments within a correlation blob obey Zimm dynamics, while the correlation blobs themselves obey Rouse dynamics. In contrast to Flory screening, however, hydrodynamic screening does not occur at very short time scales, but rather sets in only after a time scale corresponding to the onset of chain-chain collisions, which is the Zimm relaxation time of a correlation blob~\cite{Ahlrichs2001}. The establishment of this important result, along with a conclusive demonstration of the equivalence of the two screening length scales, was made possible through the development of the first hybrid mesoscale algorithm for semidilute polymer solutions that was able to account for hydrodynamic interactions, by combining molecular dynamics simulations for the polymer with a lattice Boltzmann technique for the solvent~\cite{BD1999}. There is also strong experimental support for de Gennes' hypothesis that the Flory and hydrodynamic screening length scales are the same and have the same dependence on concentration~\cite{Wiltzius1984,Zhang1999}. Appreciation of the role played by the screening of hydrodynamic interactions is an essential element of our current understanding of the dynamics of semidilute polymer solutions.

It is worth mentioning that the conventional presumption that hydrodynamic interactions are completely screened in a melt is not entirely accurate. Careful recent simulations suggest that the dynamic coupling of the motion of a single-chain with the viscoelastic matrix in which it is submerged, and which relaxes on the same time scale as the chain, gives rise to subtle corrections to Rouse dynamics~\cite{Farago2012a,Farago2012b}.

Using the concept of a correlation blob, it is possible to derive scaling laws for a number of different dynamic properties of polymer solutions in the semidilute regime, when the solvent is either under $\theta$ conditions, or under perfectly good solvent conditions~\cite{degennes,Rubinstein2003}. For instance, the polymer contribution to viscosity $\eta_{p}$, the single chain diffusivity $D$, and the longest relaxation time $\lambda_1$ can be shown to obey the following power laws
\begin{equation}
 \label{eq:sdscaling}
\frac{\eta_{p} }{ \eta_{s} } = \left( \frac{c}{c^*} \right)^{\tfrac{1}{3 \nu - 1}} \, 
; \quad \frac{D}{D_\text{Z} } = \left( \frac{c}{c^*} \right)^{- \tfrac{1 - \nu}{3 \nu - 1}} ; \quad \frac{\lambda_1}{\lambda_{1,0}} = \left( \frac{c}{c^*} \right)^{\tfrac{2 - 3\nu}{3 \nu - 1}} 
\end{equation}
where $\eta_{s}$ is the solvent viscosity, $D_\text{Z}$ is the chain diffusivity in the dilute limit, given by $D_\text{Z}  = k_\text{B}T/\eta_{s} R_F$, and $\lambda_{1,0}$ is the longest relaxation time in the dilute limit, given by $\lambda_{1,0} = \eta_{s} R_F^3/k_\text{B}T$. A number of different experimental studies have established the validity of these scaling laws, when $\nu=(1/2)$ is used for $\theta$-solvents, and $\nu = 0.588$ is used for perfectly good solvents~\cite{Rubinstein2003}.

Careful Monte Carlo simulations based on a lattice model~\cite{Paul1991} have established the occurrence of Flory screening, and validated the scaling prediction of a crossover from SAW to RW statistics on a length scale that has a concentration dependence identical with that of $\xi_c$ given above. Using mesoscale simulations which combine molecular dynamics simulations and the multiparticle collision dynamics approach, Huang et al.~\cite{Huang2010} have shown that scaling predictions in an athermal solvent for the equilibrium coil size $R$ (Eq.~(\ref{eq:Rsemidilute})), correlation blob size $\xi_c$, and the longest relaxation time $\lambda_1$ (Eq.~(\ref{eq:sdscaling})), are all satisfied in the semidilute regime. The relaxation time $\lambda_1$ was obtained by fitting an exponential function to the tail end of the decay of the mean square end-to-end distance of a chain as it relaxed to equilibrium from an initially extended state. Note that an accurate prediction of the scaling of relaxation time (a dynamic property), requires the incorporation of hydrodynamic interactions, which is made possible through the use of the multiparticle collision dynamics approach~\cite{Gompper2009}. An excellent brief summary of various approaches to treating hydrodynamic interactions in simulations, and their comparative merits can be found in the recent review article by D\"{u}nweg~\cite{Duenweg2018}.

\subsection{Double crossover \label{sec:double}}

\begin{figure}
\begin{center}
\resizebox{\linewidth}{!}{\includegraphics*{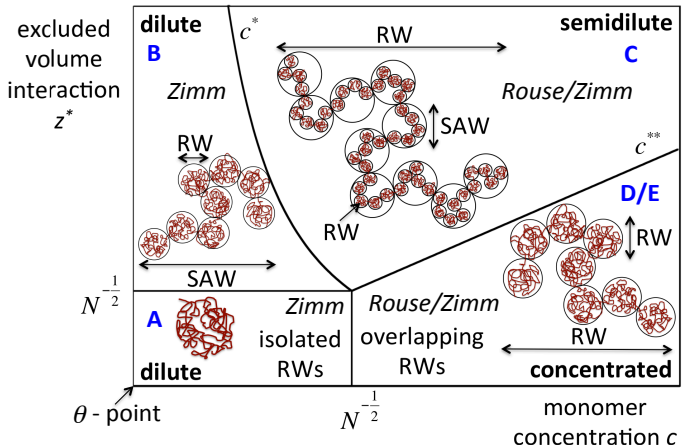}}
\end{center}
\vskip-10pt
\caption{Phase diagram of a polymer solution in the
  plane monomer concentration $c$ and excluded-volume strength
  $z^*$, as predicted by the standard blob picture. Reproduced with permission from Ref.~\cite{Jain2012a}.  }
\label{fig:unscaled_phase_diagram}
\vskip-10pt
\end{figure}

Polymer solutions are often encountered at temperatures $T$ that lie between $T_\theta$ and the athermal limit, and  simultaneously with monomer concentrations $c$ that lie in the semidilute regime, i.e., in the double crossover regime encompassing both thermal and concentration crossovers. In this case, both thermal and correlation blobs become relevant, and properties of the solution depend on a subtle interplay between the different blob length scales. For concentrations just above $(c/c^*)$, and good solvent conditions, we expect $b < \xi_T < \xi_c < R$, with chain statistics changing from RW to SAW to RW with increasing segment length. As the concentration increases beyond $c^*$, the correlation blob size $\xi_c$ decreases, until at the threshold concentration $c^{**}$, $\xi_c \approx \xi_T$. Note that in the athermal limit, this definition of $c^{**}$ coincides with that given earlier, since $ \xi_T \approx b$. Since the correlation blob is smaller than the thermal blob for $c \gtrsim c^{**}$, the chain obeys RW statistics on all length scales. 

The different regimes of behaviour discussed so far are shown schematically as a phase diagram in the space of coordinates $(z^*,c)$ in  Fig.~\ref{fig:unscaled_phase_diagram}, (reproduced here from Ref.~\cite{Jain2012a}). In particular, the double crossover regime discussed above is denoted as regime C. From the analysis carried out in Sections~\ref{sec:thermal} and~\ref{sec:conc}, we expect all universal properties of polymer solutions in regime C to depend on both the scaling variables $z$ and $(c/c^*)$.

\begin{figure}[t]  \centering
\resizebox{\linewidth}{!}{\includegraphics*{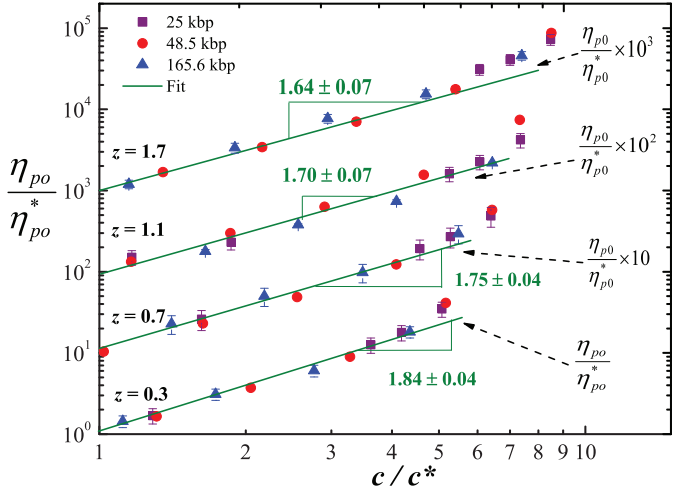}}
\vskip-10pt
\caption{Dependence of the viscosity ratio $\eta_{p0}/\eta_{p0}^{*}$
  on the scaled concentration $c/c^{*}$ in the semidilute regime, for
  25 kbp, $\lambda$ and T4 DNA, at fixed values of the solvent quality
  $z$. In order to display all the measurements on a single plot, viscosity ratios for the different values of $z$ have been multiplied by different fixed factors as indicated. Lines through the data are fits to the experimental data, with slopes and error in the fitted slope as shown. Reproduced with permission from Ref.~\cite{Pan2014a}.}
\label{fig:semidilutevis}
\vskip-10pt
\end{figure}

This prediction was tested experimentally for a linear viscoelastic property by Pan et al.~\cite{Pan2014a}, who measured the zero shear rate viscosity $\eta_{p0}$ of DNA solutions (of three different molecular weights), for a range of temperatures and concentrations in the semidilute regime. The results of their experimental measurements are shown in Fig.~\ref{fig:semidilutevis}, where $\eta_{p0}^{*}$ is the value of the zero shear rate viscosity at $c^*$. The data collapse for different molecular weights, temperatures and concentrations, when represented in terms of $z$ and $(c/c^*)$, clearly validates the universal prediction of scaling theory. Indeed, the experimental data suggests the stronger result that the viscosity ratio obeys the power law
\begin{equation}
 \label{eq:visscaling}
\frac{\eta_{p0} }{ \eta_{p0}^{*}} = \left( \frac{c}{c^*} \right)^{\tfrac{1}{3 \nu_\text{eff}(z) - 1}} 
\end{equation}
which is identical to that for athermal solvents (Eq.~(\ref{eq:sdscaling})), but with the Flory exponent $\nu$ replaced with an effective exponent $\nu_\text{eff}(z)$ that depends on the location of the solution in the thermal crossover regime, which is determined by the solvent quality $z$.

Recent single molecule experiments by Schroeder and co-workers~\cite{Hsiao2017}, which measured the relaxation of single molecules of $\lambda$-phage DNA in the semidilute unentangled concentration regime at a fixed temperature of 22 \degC, also found a power law dependence on $(c/c^*)$. Fitting the data with the exponent given in Eq.~(\ref{eq:sdscaling}) for the longest relaxation time, it was found that the slope was consistent with an effective exponent $\nu_\text{eff}(z)$ that corresponded to the value of $z$ at this temperature. Previous studies on single molecule relaxation of T4 DNA at a fixed temperature, also show power law scaling of the longest relaxation time with $(c/c^*)$, with a similar effective scaling exponent~\cite{Steinberg2009}.

Power law scaling in the double crossover regime was also observed by Jain et al.~\cite{Jain2012a}, who used a multiparticle Brownian dynamics algorithm with hydrodynamic interactions  incorporated~\cite{JainPRE}, to calculate the equilibrium size of a polymer coil, and the single chain diffusivity, under semidilute conditions. Universal predictions for the concentration dependence of these quantities was obtained by extrapolating finite chain data to the long chain limit~\cite{Jain2012a}. It was found that the ratios $(R/R^*)$, and $(D/D^*)$ obeyed power laws identical to their respective forms in an athermal solvent (Eqs.~(\ref{eq:Rsemidilute}) and~(\ref{eq:sdscaling})), with the Flory exponent $\nu$ replaced with an effective exponent  $\nu_\text{eff}(z)$. Here, $R^*$ and $D^*$ are the values of $R$ and $D$ at $c^*$. Importantly, the value of $\nu_\text{eff}$ was the same for all the scaling functions, at identical values of $z$. A concise derivation of all the relevant scaling laws in the different regimes of the phase diagram given in Fig.~\ref{fig:unscaled_phase_diagram}, is presented in the supplemental material of Jain et al.~\cite{Jain2012a}.

An interesting consequence of the scaling analysis in Ref.~\cite{Jain2012a} is that the different crossover scaling functions are not independent of each other, but rather can be calculated once one of them is known, since they satisfy the following relations in the regimes C, D and E of the phase diagram, all of which lie above the overlap concentration $c^*$
\begin{equation}
\label{usc1}
\left(\frac{c}{  c^*}\right)^{\tfrac{1}{4}} \left(\frac{R}{  R^*}\right) \left(\frac{D}{  D^*}\right)^{\tfrac{1}{4}} =  1  \,; \quad
 \left(\frac{c}{  c^*}\right)^{\tfrac{1}{3}} \left(\frac{R}{  R^*}\right) \left(\frac{\eta_{p0} }{ \eta_{p0}^{*}}\right)^{-\tfrac{1}{6}} =  1 
\end{equation}
The significance of this result is that there is essentially only a single universal crossover scaling function in these three regimes. The validity of the first expression in Eq.~(\ref{usc1}) has been verified by  Jain et al.~\cite{Jain2012a} by carrying out Brownian dynamics simulations followed by extrapolation of finite chain data to the long chain limit. They also verified the second  expression in  Eq.~(\ref{usc1}) by combining their simulation results for the ratio  $(R/R^*)$ with the experimental results of Pan et al.~\cite{Pan2014a} for the ratio ${\eta_{p0} }/{ \eta_{p0}^{*}}$. A complete experimental validation of the expressions in Eq.~(\ref{usc1}) has, however, not been carried out so far. 

\section{Universal behaviour far from equilibrium \label{sec:nonequil}}

\subsection{Shear flow \label{sec:shear}}

\subsubsection{Dilute solutions  \label{sec:dilshear}}

The behaviour of dilute polymer solutions subjected to shear flow has been studied for several decades now, both experimentally and theoretically. In this section, we take stock of these results, and examine the question of whether the dependence of the polymer contribution to viscosity $\eta_{p}$ on shear rate $\dot \gamma$, can be considered to be universal. 

A common pattern observed in all experimental measurements of shear viscosity is the existence of a constant Newtonian plateau at low shear rates, termed the zero shear rate viscosity ($\eta_{p0}$), followed by shear thinning at higher shear rates with a power law decay at sufficiently high shear rates~\cite{birdetal1,Noda:1968gt,Kotaka:1966cl,Suzuki:1969ec,Mochimaru:1983do,Hua2006787,Hur2001421,Teixeira:2005p1058,Schroeder20051967}. Some experimental measurements have also observed a second Newtonian plateau at even higher shear rates~\cite{Noda:1968gt,Hua2006787}. In spite of this overall similarity in the observed behaviour, there is wide disparity in the reported power law exponent in the shear thinning regime, and in the measured influence of temperature and molecular weight on the shear rate for the onset of shear thinning. On the other hand, several studies have shown that when the viscosity versus shear rate data is plotted as the viscosity ratio  ${\eta_{p} }/{ \eta_{p0}}$ versus $\lambda \dot \gamma$, where $\lambda$ is an appropriately chosen relaxation time, master plots can be obtained, independent of a number of parameters such as the temperature, concentration and polymer-solvent system~\cite{Kotaka:1966cl,Suzuki:1969ec,Mochimaru:1983do,Hua2006787}. The construction of these master plots, however, is not carried out in a systematic manner, with no attempt to achieve consistency with the equilibrium framework for examining universal behaviour, which (as we have seen above) requires the representation of the dependence of properties on $T$, $M$ and $c$ to be in terms of $z$ and $c/c^*$. 

An important aspect of universal properties is that they are independent of chain length. This is particularly pertinent when polymer solutions are subjected to flow since a dependence on chain length implies that the flow is strong enough for the finite length of the chain to be revealed, and for the molecular details of the polymer chain to become relevant, leading to non-universal behaviour. In the case of shear flow, theoretical predictions have shown that the finite length of a chain leads to shear thinning~\cite{birdetal2}. On the other hand, shear thinning can also be caused by other phenomena such as excluded volume and hydrodynamic interactions~\cite{wedgewood:jnnfm-88,Ottinger:1989up,ott90,zylka:jcp-91,Prakash1997,Aust19995660,prakash:jor-02,Jendrejack20027752,kumar:jcp-04,PraRavi04,Winkler2006,Ryder:2006ig,Winkler2010,Dalal2012,Dalal2014}, that are not related to chain length. These observations clearly suggest that there are both universal and non-universal contributions to shear thinning, and that separating their influence while interpreting experimental measurements can be tricky. These issues have been examined recently by Pan et al.~\cite{Pan2018}, who have estimated the values of $z$ and $(c/c^*)$ for a large volume of data reported previously for dilute polystyrene solutions subjected to shear flow, and examined if there was any data collapse when ${\eta_{p} }/{ \eta_{p0}}$ is plotted versus $\lambda_\text{Z} \dot \gamma$, where $\lambda_Z = {M [\eta]_0 \eta_s}/{N_A k_B T}$ is a large scale relaxation time based on the zero shear rate value of the intrinsic viscosity $[\eta]_0$. They also measured the shear viscosities of solutions of three different molecular weights of DNA, across a range of temperatures and concentrations in the dilute regime. Estimates of $z$ and $c/c^*$ for DNA solutions, reported earlier in Ref.~\cite{Pan2014a}, were used to interpret their observations. They found that whilst the data for DNA solutions at different temperatures and concentrations in the dilute regime collapsed onto master plots for a fixed molecular weight, the data for different molecular weights were independent of each other, and there was no data collapse. Essentially, the onset of shear thinning was postponed to higher and higher values of $\lambda_\text{Z} \dot \gamma$ with increasing molecular weight. In contrast, the data for polystyrene solutions measured by other researchers previously, showed a significant variation in behaviour in terms of the onset of shear thinning, the shear thinning exponent, and the existence of a terminal plateau, in spite of the interpretation of the data in terms of the scaling variables $z$ and $c/c^*$. The significant differences in the behaviour of DNA and polystyrene solutions that was observed raises the question of whether these variations are due to the distinctness of DNA and polystyrene on a molecular scale, and calls for more systematic studies to settle this question. 

The results of analytical theories and simulations are not much clearer in their conclusions with regard to the existence of universality. A comprehensive discussion of the current state of affairs is given in Ref.~\cite{Pan2018}. Here, the key aspects are briefly summarised. The results of various theoretical studies indicate that it is important to include both excluded volume and hydrodynamic interactions in order to obtain at least qualitative similarity with experimental observations. Several analytical and computational studies based on bead-spring chain models have obtained universal predictions for the dependence of the viscosity ratio on the non-dimensional shear rate when the springs are Hookean, implying that the chains are infinitely extensible~\cite{Ottinger:1989up,ott90,Prakash1997,prakash:jor-02,kumar:jcp-04}. When finitely extensible springs are used to account for the finite contour length of real polymers, universal behaviour persists until shear rates at which the finite length of the chain becomes important~\cite{kisbaugh:jnnfm-90,Prabhakar2006}. The most significant issue, however, is that when hydrodynamic interactions are incorporated, theoretical bead-spring chain models predict shear thickening over an extended interval of intermediate shear rates, which has not been observed experimentally. The situation is further complicated when simulations with bead-rod chains are considered, a large majority of which are for free-draining chains~\cite{Doyle1997251,Liu19895826}. Only a few studies have included excluded volume and hydrodynamic interactions~\cite{Petera19997614,Moghani2017}. Bead-rod simulations do not predict shear thickening, and they also predict a shear thinning power law exponent which is different from that predicted by bead-spring chain models. While a second Newtonian plateau is predicted by the former, it is not by the latter. In spite of the many decades of work on this problem, there has not been an attempt to carry out a systematic quantitative comparison of theoretical predictions with experimental results, beyond obtaining qualitative similarity. The origin of the difference between predictions by the different models for the polymer are also not entirely clear. A proper resolution of this issue would be helpful to understand the conditions under which local chain details become important. Clearly, significant progress is required on the theoretical front before one can understand the origin of the wide variability in experimental observations.

\begin{figure}[tbp]  \centering
\resizebox{\linewidth}{!}{\includegraphics*{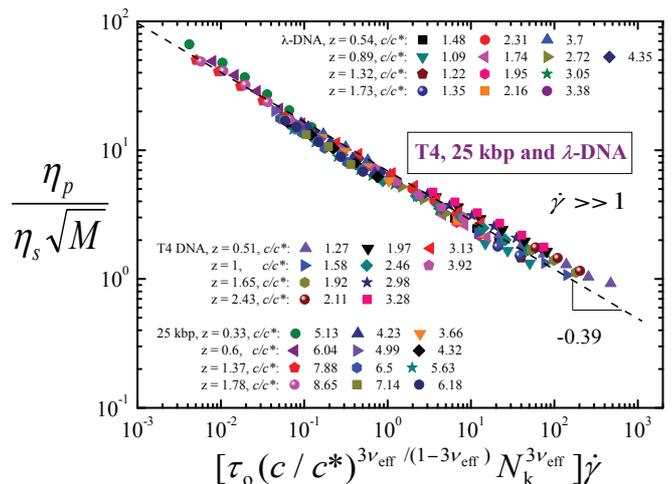}}
\vskip-10pt
\caption{Data collapse for semidilute DNA solutions, across molecular weights, concentrations and solvent quality, when the polymer contribution to viscosity \etap\ is scaled by $\etas \sqrt{M}$. Here, $\tau_0$ is the monomer relaxation time, and $N_\text{k}$ is the number of Kuhn steps in a chain. Reproduced with permission from Ref.~\cite{Pan2018}}
\label{fig:Colby3MW}
\vskip-10pt
\end{figure}

\subsubsection{Semidilute solutions  \label{sec:semidilshear}}

The rheological behaviour of semidilute polymer solutions has not been examined as extensively as that of dilute polymer solutions, and experimental and theoretical studies are comparatively sparse~\cite{Hur2001421,Heo20051117,Heo20088903,Pan2014a,Hsiao2017,Prabhakar2017,Stoltz2006,Huang2010,Jain2012a,JainPRE,Saadat2015b,Sasmal2017}. The measurements on semidilute solutions of DNA in shear flow by Hur et al.~\cite{Hur2001421} revealed that at sufficiently high shear rates, the polymer contribution to the viscosity had a power law scaling with Weissenberg number, $\eta_p \approx W\!i^{-0.5}$, for scaled concentrations $c/c^* = 1, 3, \text{and} \, 6$, with  $W\!i$ defined by $W\!i = \lambda_1\dot \gamma$. Heo and Larson~\cite{Heo20088903} showed for semidilute solutions of polystyrene in tricresyl phosphate, in the concentration range $c_\text{e} < c < c^{**}$, that data for different molecular weights could be collapsed onto a universal curve when the scaled viscosity, $\eta_\text{p}/\eta_s (c [\eta]_0)^{1/(3\nu -1)}$, was plotted versus $\lambda_\text{e} \dot \gamma$, where  $c_\text{e}$ is the entanglement concentration, and $\lambda_\text{e}$ is the Rouse time of an entanglement strand. However, curves for different $c/c_\text{e}$ did not collapse on top of each other. 

The first attempts to  computationally estimate viscometric functions for semidilute unentangled polymer solutions in simple shear flow with bead-spring chain simulations were carried out by Stoltz et al.~\cite{Stoltz2006} (using Brownian dynamics), and Huang et al.~\cite{Huang2010} (using multiparticle collision dynamics). By plotting $\eta_\text{p}/\eta_{\text{p}0}$  versus $W\!i$ for fixed chain length, Huang et al.~\cite{Huang2010} showed that data for different values of $(c /c^*)$ collapsed onto a master plot. However, a weak dependence on chain length was observed when data for chains with 50 and 250 beads were plotted together. For large values of Weissenberg number, regardless of concentration and chain length, they observe power law shear thinning, with an exponent equal to $-0.45$, roughly in agreement with the experimental observations of Hur et al.~\cite{Hur2001421}.   

Pan et al.~\cite{Pan2018} have recently reported viscosity versus shear rate measurements for solutions of three different molecular weights of DNA, and for solutions of polystyrene in dioctyl phthalate (DOP) of two different molecular weights, across a range of temperatures and concentrations in the unentangled semidilute regime. They found that when $\etap / \etapo$ is plotted against $\lambda_\eta \dot \gamma$, where $\lambda_\eta = {M \,  \eta_{\text{p}0}}/{c N_A k_B T}$, the various curves at \textit{different} temperatures, but at the \textit{same} concentration collapsed on top of each other, both for the DNA and the polystyrene solutions. However, unlike the observations in simulations~\cite{Huang2010}, the data did not collapse for different values of $(c/c^*)$\ or molecular weight. Nor was the slope in the power-law region close to $- 0.5$, as observed in the experiments of Hur et al.~\cite{Hur2001421} and in the simulations of Stoltz et al.~\cite{Stoltz2006} and Huang et al.~\cite{Huang2010}. Rather, a significant power law regime over several decades of $\leta \dot{\gamma}$ was observed in the case of DNA, with the magnitude of the slope increasing with increasing concentration. Note that the relaxation time, $\lambda_\eta$,  defined in terms of the polymer contribution to the zero shear rate viscosity has the same dependence on concentration in the semidilute regime as the longest relaxation time $\lambda_1$, and their ratio is a universal constant~\cite{Pan2014a,Hsiao2017}. The reason for the lack of agreement between the experimental observations on DNA and polystyrene by Pan et al.~\cite{Pan2018}, with the predictions of simulations~\cite{Huang2010} suggesting data collapse across different values of (\ccs), is currently not clear. 

By developing a scaling theory for semidilute unentangled polymer solutions that pictures a polymer chain as a blob pole of Pincus blobs~\cite{Pincus1976} in the presence of flow, Pan et al.~\cite{Pan2018} argue that rather than using  $W\!i$, which is based on a large scale relaxation time, the definition of the Weissenberg number should be based on the relaxation time of a single correlation blob. As can be seen from Fig.~\ref{fig:Colby3MW}, the use of such a Weissenberg number leads to a remarkable data collapse at high shear rates, independent of solvent quality, concentration and molecular weight. A verification of this experimental observation with computer simulations is yet to be carried out.

\subsection{Extensional flow  \label{sec:ext}}

The most commonly measured properties of polymer solutions in extensional flows are the elongational viscosity (which is a bulk rheological property typically measured in uniaxial extensional flows~\cite{McKinley2002}), and the stretch of individual molecules (which is the projected extent of a molecule in the flow direction, typically measured in planar extensional flows in a cross-slot cell~\cite{Schroeder2018}). Both these measurements have proven to be invaluable for the testing and validation of molecular theories for polymer solutions, and there is an extensive literature in the field that spans the couple of decades since the mid-nineties when these measurements first became possible~\cite{Larson2005,Shaqfeh2005,Schroeder2018}. Here, we focus our attention on the universal aspects of these properties and their prediction. 

Polymer coils in extensional flows undergo a coil to stretch transition if the elongation rate $\dot \epsilon$ is beyond a critical threshold value, and if the Hencky strain, $\epsilon = \dot \epsilon t$, is sufficiently high, where $t$ is the time since the inception of flow. For values of $\dot \epsilon$ and $\epsilon$ where the polymer remains coiled, we can expect, because of scale invariance, that measured properties will be universal for sufficiently long polymers. Beyond the critical elongation rate, however, when chains are in a stretched state, the finiteness of the chain length begins to play a crucial role in determining the solution's properties, and one anticipates that at sufficiently high strain rates, local details of the polymer will become important, leading to non-universal behaviour. It is possible to obtain an intuitive understanding of when local details might become important by invoking the concept of Pincus blobs~\cite{Pincus1976}. For sufficiently high values of $\dot \epsilon$ and $\epsilon$, the conformation of a chain breaks up into a sequence of Pincus blobs of size $\xi_\text{P}$, which is the length scale at which the stretching energy in a chain segment becomes of order $k_{B}T$. Under these conditions, the conformation of the chain is rodlike on length scales above $\xi_\text{P}$, but for smaller length scales, chain segments have equilibrium conformations. Essentially, the flow does not ``penetrate'' the chain on length scales below $\xi_\text{P}$, and equilibrium conditions apply on these short length scales. Provided the conformations of chains with different local properties are the same on length scales large compared to the Pincus blob, their long time and large scale behaviour will be identical, and macroscopic properties remain independent of local details. At steady state, the transition length scale $\xi_\text{P}$ at which a chain switches from its equilibrium conformation to a deformed conformation depends on the Weissenberg number $W\!i$, and it can be expected that for values of $W\!i$ beyond the point at which  $\xi_\text{P} \approx b$, the chain is fully stretched, and local details will play a role in determining the solution's properties. 

There have been no experimental observations so far that have attempted to examine the universal behaviour of polymer solutions in extensional flow (apart from a brief examination of the influence of molecular weight on the uniaxial elongational viscosity of polystyrene solutions at a fixed Weissenberg number~\cite{gupta00}), as they are largely focussed on the properties of particular polymer-solvent systems. The discussion of universality in this section is,  consequently, mainly focussed on the results of Brownian dynamics simulations that have addressed this issue. 

\subsubsection{Dilute solutions  \label{sec:dilext}}

\begin{figure}[tbp]  \centering
\resizebox{\linewidth}{!}{\includegraphics*{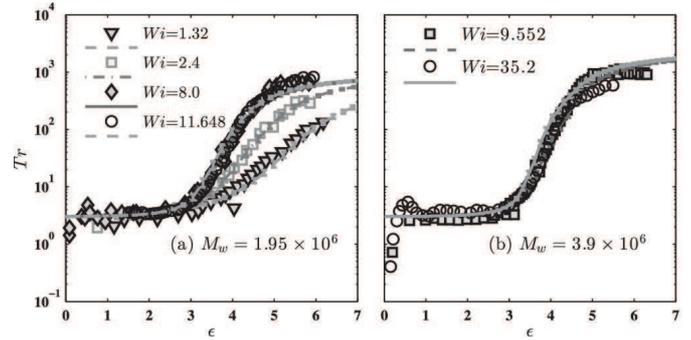}}
\caption{Comparison of the dependence of the Trouton ratio $T\!r$ on the Hencky strain $\epsilon$, predicted by Saadat and Khomami~\cite{Saadat2015} using successive fine-graining, with the experimental observations of Gupta et al.~\cite{gupta00}. Symbols are experiments and lines are simulation predictions. Reproduced with permission from Ref.~\cite{Saadat2015}.}
\label{fig:saadat}
\end{figure}

The key experiments that have motivated theoretical developments in modelling the extensional behaviour of dilute polymer solutions are the measurements of the uniaxial elongational viscosity of polystyrene~\cite{gupta00} and DNA solutions~\cite{Sunthar2005b}, and single molecule experiments on stained DNA molecules~\cite{Schroeder2018}. The qualitative aspects of theoretical predictions will be discussed first before reviewing attempts at quantitatively predicting experimental data.  

By considering a bead-spring chain model with Hookean springs, and with excluded volume and hydrodynamic interactions incorporated, Sunthar and Prakash~\cite{Sunthar2005} used Brownian dynamics simulations to calculate the expansion ratio, $E$, defined by
\begin{equation}
\label{ExpRatio}
E = \frac{\bar{X}}{\bar{X}_\text{eq}}
\end{equation}
where $\bar{X}$ is the nondimensional mean stretch of the molecule in the flow direction, and its equilibrium value is denoted by $\bar{X}_\text{eq}$. By carrying out an extrapolation of finite chain data to the long chain limit, at various values of the Hencky strain, they demonstrated that the transient stretch experienced by an infinitely long polymer chain is universal in the nondraining limit, since in this limit $E$ is independent of the choice of the hydrodynamic interaction parameter $h^*$ (which is a measure of the nondimensional bead radius), and the range of the excluded volume potential $d^*$.  The expansion ratio was shown to increase without bound for increasing values of Hencky strain, never reaching a steady-state value since the infinite polymer coil is unravelled continuously, and it does not undergo a coil-stretch transition due to the existence of infinite length scales. 

The situation is very different for finite chains since the transient expansion ratio $E$ levels off to an asymptotic steady state value at large strains. At a threshold value of strain, the finite character of the chain leads to the existence of a point of inflection on the stretch versus strain curve where the slope stops increasing and starts to decrease. The existence of universality in this case can be examined with the help of the successive fine graining procedure developed by Prakash and coworkers~\cite{Sunthar2005,prabhakar04,Pham}. Essentially, in this method, bead-spring chains models with finitely extensible springs are used, and rather than extrapolating finite chain data to the limit $N \rightarrow \infty$, the extrapolation is carried out to the limit $N_k$, where $N_k$ is the number of Kuhn steps in the underlying chain. The details of the method are fairly subtle, and given in detail in Refs.~\cite{Sunthar2005} and~\cite{Sasmal2017}. With the successive fine graining technique, Sunthar and Prakash~\cite{Sunthar2005anziam} have shown for an  unraveling infinite length polymer (i.e., for a bead-spring chain model with Hookean springs), that the $E$ versus $\epsilon$ curve acts as an envelope for all finite chains. Essentially, for a given value of the Weissenberg number, the strain at which a finite chain curve departs from the Hookean curve, depends on the value of $N_k$. For increasing values of $N_k$, the point of departure occurs at increasing values of strain. For all values of strain where extrapolated values of $E$ predicted by bead-spring chains models with finitely extensible springs coincide with the values for Hookean bead-spring chains, Sunthar and Prakash~\cite{Sunthar2005} found that the choice of spring force law is irrelevant. This is in line with the universality of the result, which is independent of local details such as the particular choice of the spring force law. However, at sufficiently large value of Weissenberg number and Hencky strain, curves for different spring force laws, and microscopic parameters such as the hydrodynamic interaction parameter, begin to come apart indicating the loss of universality, and the importance of local details.

The successive fine graining procedure has proved to be a powerful methodology for confirming if experimental measurements can be captured accurately by simulation predictions, since  model predictions with this technique are made without fitting any parameters. In particular, it  enables an objective assessment of molecular theories, and verifies if all the relevant physics have been adequately taken into account. Quantitatively accurate predictions of the expansion ratio for dilute $\lambda$-phage DNA solutions measured by Smith and Chu~\cite{Smith&Chu} in a cross slot cell have been obtained by Sunthar and Prakash~\cite{Sunthar2005} for a range of Weissenberg numbers and Hencky strains with the help of this method. Similarly, experimental measurements by Gupta et al.~\cite{gupta00} of the uniaxial elongational viscosity of a near-$\theta$ dilute solution of polystyrene (molecular weight $1.95\times10^6$) has been predicted accurately by Prabhakar et al.~\cite{prabhakar04}, and measurements of the uniaxial elongational viscosity for a dilute solution of $\lambda$-phage DNA molecules in a good solvent have been predicted by Sunthar et al.~\cite{Sunthar2005b}. By developing a computationally efficient Brownian dynamics algorithm, which enables them to simulate much longer chains than those that were simulated in Ref.~\cite{prabhakar04}, Saadat and Khomami~\cite{Saadat2015} have recently used successive fine-graining to obtain accurate predictions of the Trouton ratio (which is a ratio of the uniaxial elongational viscosity to the zero shear rate viscosity) for solutions of Polystyrene chains with molecular weights up to $3.9\times10^6$ measured by Gupta et al.~\cite{gupta00}, as shown in Fig.~\ref{fig:saadat}. However, elongational viscosity data at high strains, for chains with molecular weights of $10.2\times10^6$ and $20\times10^6$, were not predicted as accurately, presumably due to errors in extrapolation from the lack of finite chain data for sufficiently long chains. The successive fine graining procedure has also recently been used to obtain excellent agreement between experimental measurements and Brownian dynamics simulations of wall tethered DNA molecules in shear flow~\cite{Lin2017}.

\subsubsection{Semidilute solutions  \label{sec:semidilext}}

\begin{figure}[t]
\centering
\resizebox{\linewidth}{!}{\includegraphics*{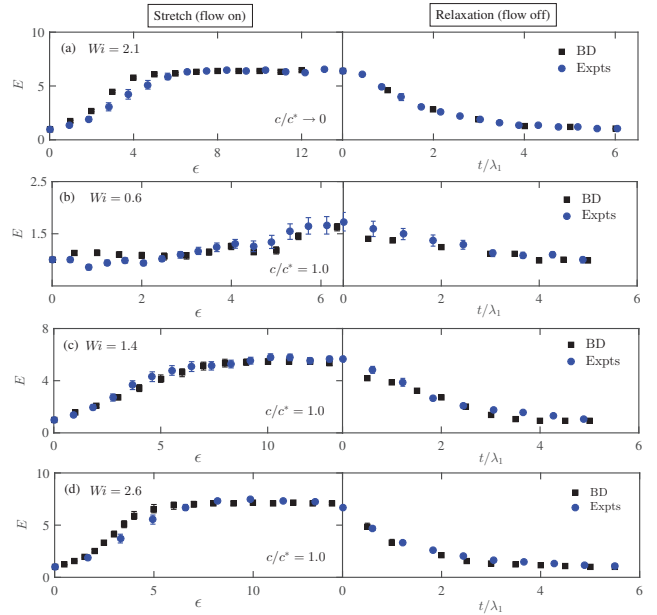}}
\caption{ \footnotesize Comparison of the expansion factor $E= {\bar{X}} /{\bar{X}}_\text{eq}$ predicted by successive fine-graining with the experimental observations of Hsiao et al.~\cite{Hsiao2017}. The top panel corresponds to a dilute solution at $W\!i = 2.1$. The remaining panels correspond to semidilute solutions at $c/c^* = 1$, and $W\!i = \{ 0.6, 1.4, 2.6 \}$, respectively. Reproduced with permission from Ref.~\cite{Sasmal2017}. \label{StretchRelaxation}}
\vskip-10pt
\end{figure}

The quantification of the extensional viscosity of semidilute polymer solutions has only recently been made possible due to developments in solution preparation, and in measurement techniques~\cite{Pan2013,Dinic2017,Dinic2017b,Prabhakar2017,Hsiao2017b}. The universal aspects of the behaviour, have however, not yet been examined experimentally or theoretically because the field is in an early stage of development. On the other hand, the recent parameter-free prediction of the stretch and relaxation of individual molecules in a semidilute solution of $\lambda$-phage DNA molecules, suggests that the behaviour might be universal~\cite{Hsiao2017,Sasmal2017}. It should be noted that, as in the case of the measurement of the extensional viscosity of dilute polymer solutions, universality has not been established unequivocally by experiments. By developing the \emph{Stokes trap}~\cite{Shenoy2016}, which is an advanced version of the cross-slot cell, Hsiao et al.~\cite{Hsiao2017} have examined a semidilute solution of $\lambda$-phage DNA molecules subjected to planar extensional flow. In particular, they have studied the response of individual chains to step-strain deformation followed by cessation of flow, and as a result, captured both chain stretch and relaxation in a single experiment. The results of their experimental measurements of the expansion ratio $E$ at various Weissenberg numbers, in a dilute and a semidilute solution, are shown in Fig.~\ref{StretchRelaxation}.

By implementing the Kraynik-Reinelt periodic boundary conditions for mixed flows~\cite{Kraynik,Hunt}, Prakash and co-workers have recently developed an optimised multi-particle Brownian dynamics algorithm that can simulate arbitrary planar mixed shear and extensional flows of semidilute polymer solutions~\cite{JainPRE,JainCES}. With the help of this algorithm, they have carried out the successive fine-graining technique in order to compare simulation predictions~\cite{Sasmal2017} with the experimental measurements of Hsiao et al.~\cite{Hsiao2017}. They find that finite chain data extrapolated to the number of Kuhn steps in a $\lambda$-phage DNA molecule, is independent of the choice of simulation parameters (such as $h^*$ and $d^*$) at all the experimentally measured values of strain in the stretch phase, and time in the relaxation phase, suggesting the universal nature of their predictions. The quantitatively accurate prediction of the experimental observations of $E$, in both the stretching and relaxation phases, achieved by this procedure is displayed in Fig.~\ref{StretchRelaxation}. As discussed earlier, it can be anticipated that at sufficiently high Weissenberg numbers and Hencky strains, local details at the molecular scale will become relevant, and simulation predictions will not be parameter free. By developing a scaling argument based on the relative magnitudes of Pincus, thermal and correlation blobs, an estimate of the Weissenberg number at which the successive fine-graining technique can be expected to breakdown has been given in Ref.~\cite{Sasmal2017}.

It is worth mentioning here that, recently, an important generalisation of the original Kraynik-Reinelt boundary conditions (that are only applicable to planar elongational flow and general planar mixed flow) has been achieved, which enables one to carry out arbitrarily long simulations for a larger class of three-dimensional flows, including uniaxial and biaxial flows~\cite{Dobson2014, Hunt2015}. This new methodology has been used to study the uniaxial extensional flow of polymer melts, simulated at various levels of coarse-graining~\cite{Nicholson2016,OConnor2018,Murashima2018}.

\section{Conclusions and future outlook \label{sec:conclu}}

Recent experimental observations, scaling theories, and computer simulations that have uncovered hitherto unexplored universal equilibrium and rheological properties of dilute and semidilute solutions of linear, flexible, neutral polymers, have been surveyed. These include: (i) the universal swelling of the viscosity radius in dilute DNA solutions, and its theoretical prediction with Brownian dynamics simulations, (ii) the demonstration that there is only one universal equilibrium crossover scaling function in the double crossover regime driven by temperature and concentration, and that scaling functions in this regime have the same functional form as in athermal solvents, albeit with an effective scaling exponent instead of the Flory exponent, (iii) the experimental validation of the latter scaling prediction with measurements carried out on the zero shear rate viscosity of semidilute DNA solutions and its theoretical verification by Brownian dynamics simulations, (iv) the collapse of shear thinning data for DNA solutions onto master plots independent of temperature and concentration (but not of molecular weight) in the dilute regime, (v) the quantitative parameter-free prediction of the elongational viscosity of dilute high molecular weight polystyrene solutions with the help of the successive fine-graining technique, (vi) the collapse of the high shear rate shear thinning data for semidilute DNA solutions when the shear rate is non-dimensionalised with the relaxation time of a correlation blob, and finally (vii) the measurement of the stretch and relaxation of individual DNA molecules in semidilute solutions subjected to a step-strain deformation followed by the cessation of flow in a cross-slot cell, and the quantitative parameter-free prediction of the experimental measurements with the help of the successive fine-graining technique. 

Several unresolved issues have been identified in the course of this review, which need to be clarified and worked on in the future in order to achieve a more thorough understanding of universal polymer solution behaviour. Experimentally these include: (i) the validation of the existence of only a single universal crossover scaling function in the double crossover regime, (ii) a clear identification of the different phenomenon contributing to shear thinning in dilute polymer solutions, and a careful exploration of whether shear thickening exists at intermediate shear rates, (iii) the validation of the universality of the transient uniaxial elongational viscosity of dilute polymer solutions suggested by simulations, for sufficiently long chains, and (iv) a similar validation of the universality of the transient expansion ratio of individual molecules in semidilute solutions subjected to planar elongational flow in a cross slot cell. The use of scaling arguments that suggest that polymer chains with a similar number of thermal, correlation and Pincus blobs behave similarly in a flow field, can prove useful in the design of such experiments. Future directions for computer simulations include: (i) the verification of the universality of small amplitude oscillatory material functions in good solvents, (ii) the confirmation of the universal double crossover scaling of the zero shear rate viscosity, as observed experimentally, (iii) a careful exploration of differences between predictions by bead-rod and bead-spring chain models for dilute polymer solutions in shear flow and a clear identification of the sources of the differences, (iv) an investigation of the structure of polymer chains at different length scales in a flow field, and verification of the existence of different blob scaling regimes, (v) demonstration of the collapse of shear thinning data for semidilute solutions when the Weissenberg number is defined in terms of the relaxation time for a single correlation blob, and (vi) simulation of the elongational viscosity of semidilute polymer solutions, and verification of the existence of universality for sufficiently long chains. The capacity to simulate long polymer chains, and large system sizes with many polymer chains (when periodic boundary conditions are used), is essential for many of the future tasks that have been outlined here. Encouragingly, as mentioned earlier, there have been many recent developments in simulation methodologies for dilute and semidilute solutions, that suggest that such simulations are indeed now possible, and no longer pose an insurmountable obstacle. 

\section{Acknowledgements}
This research was supported under the Australian Research Council's Discovery
Projects funding scheme (project number DP120101322). I would also like to acknowledge the many enlightening discussions I have had with Burkhard D\"{u}nweg about the subtleties of scaling laws in Polymer Physics.

\newcommand{\noop}[1]{}

\end{document}